\documentclass[preprint,aps,onecolumn]{revtex4-2}

\usepackage{graphicx,amsmath,amssymb,color}

\usepackage{amsfonts}
\usepackage{mathrsfs}
\usepackage{mathtools}
\usepackage{braket}

\begin{document}

\title{Single-pair measurement of the Bell parameter}

\author{
S. Virz\`{i},$^1$ E. Rebufello,$^{1}$ F. Atzori,$^{1,2}$ A. Avella,$^{1}$ F. Piacentini,$^{1\ast}$ R. Lussana,$^{3}$, I. Cusini,$^{3}$ F. Madonini,$^{3}$ F. Villa,$^{3}$ M.~Gramegna,$^{1}$ E. Cohen,$^{4}$ I. P. Degiovanni,$^{1}$ M. Genovese$^{1}$}
\affiliation{$^{1}$INRIM, Strada delle Cacce 91, I-10135 Torino, Italy}
\affiliation{$^{2}$Politecnico di Torino, Corso Duca degli Abruzzi 24, I-10129 Torino, Italy}
\affiliation{$^{3}$Politecnico di Milano, Dipartimento di Elettronica, Informazione e Bioingegneria, Piazza Leonardo da Vinci 32, 20133 Milano, Italy}
\affiliation{$^{4}$Faculty of Engineering and the Institute of Nanotechnology and Advanced Materials, Bar Ilan University, Ramat Gan, Israel}
\affiliation{$^\ast$To whom correspondence should be addressed; E-mail: f.piacentini@inrim.it}

\date{}

\baselineskip24pt

\begin{abstract}
Bell inequalities are one of the cornerstones of quantum foundations and fundamental tools for quantum technologies.
Recently, the scientific community worldwide has put a lot of effort towards them, which culminated with loophole-free experiments. 
Nonetheless, none of the experimental tests so far was able to extract information on the full inequality from each entangled pair, since the wave function collapse forbids performing, on the same quantum state, all the measurements needed for evaluating the entire Bell parameter.
We present here the first single-pair Bell inequality test, able to obtain a Bell parameter value for every entangled pair detected.
This is made possible by exploiting sequential weak measurements, allowing to measure non-commuting observables in sequence on the same state, on each entangled particle.
Such an approach not only grants unprecedented measurement capability, but also removes the need to choose between different measurement bases, intrinsically eliminating the freedom-of-choice loophole and stretching the concept of counterfactual-definiteness, since it allows measuring in the otherwise not-chosen bases.
We also demonstrate how, after the Bell parameter measurement, the pair under test still presents a noteworthy amount of entanglement, providing evidence of the absence of (complete) wave function collapse and allowing to exploit this quantum resource for further protocols.
\end{abstract}
\maketitle

\section{Introduction}

Since their formulation \cite{bel65}, Bell inequalities have been one of the pillars of quantum foundations and quantum information, representing also a tool of the utmost relevance for quantum technologies, e.g. for quantum communication \cite{eke91}.
Their huge relevance stems from the fact that they allow testing quantum mechanics versus not just a single alternative theory, but a whole class of them.
In these (classical) theories, called local hidden variable theories (LHVTs) \cite{gen05}, all measurement outcomes are pre-determined by some hidden variables, and no superluminal signaling is allowed.
A long path, awarded by the 2022 Nobel prize, has gone along \cite{bru13,gen05} since their introduction, finally reaching conclusive Bell inequality violations \cite{hen15,giu15,sha15}.
In the middle, one can recall the very first tests \cite{fre72,kas75,cla76} and the first experiment with space-like separated measurements \cite{asp82}.
However, none of the experiments run so far was able to extract information on the entire Bell inequality from every entangled pair measured, since the wavefunction collapse does not allow performing, on the same quantum state, all the measurements needed for inferring the whole Bell parameter at once.
This means that the experimenters performing the test must (randomly) choose, pair by pair, the measurement basis for each of the two entangled particles.
In order to avoid that these choices open loopholes in the test, they need be independent (``freedom-of-choice'' loophole) and no communication should be allowed between these measurements (``locality'' loophole).\\
Here we illustrate an actual paradigm shift in this perspective, since we experimentally demonstrate the possibility of extracting information on the full Bell parameter from each entangled pair measured and, at the same time, still conserving most of the pair entanglement after the measurement.
This is made possible by implementing two weak measurements \cite{aha88,rit91,kof12,tam13} in sequence \cite{mit07,the16,pia16seq,kim18,fol21} on each particle forming the entangled pair, allowing to perform simultaneously all the measurements needed for the Bell inequality test at the cost of inducing some minor decoherence in the entangled state instead of causing the complete collapse of its wave function.\\
Weak measurements have already represented a tool for quantum foundations experiments, e.g. in connection with quantum contextuality \cite{pia16pus,wae17,cim20}, nonlocality \cite{mah16,hu18}, Leggett-Garg inequalities \cite{gog11,ave17} and the Einstein-Podolski-Rosen paradox \cite{cal20}.
Our result represents a breakthrough application of them, since, for the first time, they allow evaluating the entire Bell parameter individually from each entangled pair detected (although with a large uncertainty, typical of weak measurements).\\
By simultaneously performing measurements on all the polarization bases needed for the Bell inequality evaluation, our protocol inherently gets rid of the freedom-of-choice loophole, since both Alice and Bob no longer need to randomly choose a specific measurement basis for each entangled pair produced.
Furthermore, it is noteworthy that, thanks to the weak interaction scheme exploited, the entangled state does not collapse after the Bell parameter measurement, allowing to make use of the residual entanglement within the state as a quantum resource for further experiments or, eventually, quantum information protocols.\\


\section{Experimental realization}

\begin{figure}
\centering
\includegraphics[scale=0.6]{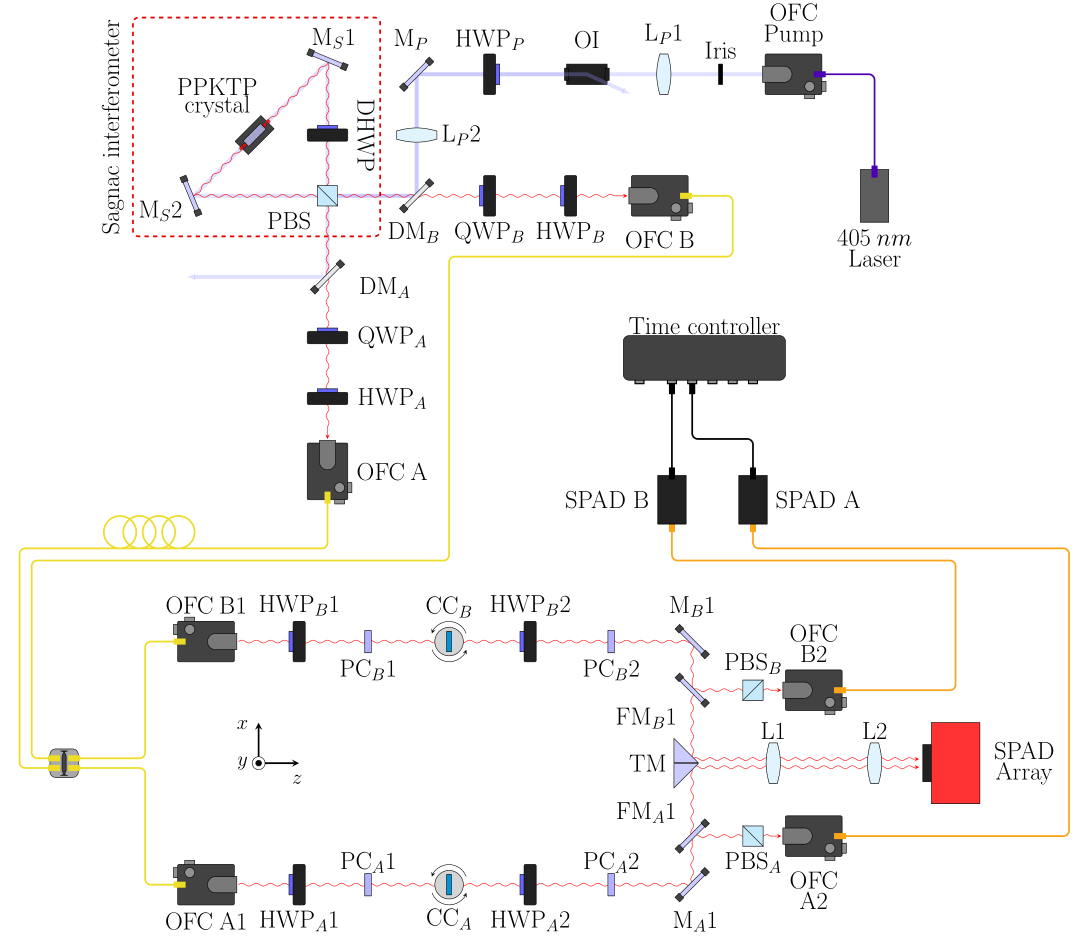}
\caption{Representation of our experimental setup. Polarization-entangled photon photons are generated in a Sagnac interferometer, fiber coupled and then collimated in a Gaussian spatial mode. Then, each of them undergoes two weak measurements in sequence, each realized by a pair of birefringent crystals preceded by a half-wave plate, before impinging on a $24\times24$ SPAD array detecting the photon pairs with 2D spatial resolution. OFC: Optical fiber coupler; L: lens; OI: Optical isolator; HWP: Half-waveplate; QWP: Quarter-waveplate; DHWP: Dual half-waveplate; DM: Dichroic mirror; M: Mirror; PBS: Polarizing beam splitter; PC: Principal crystal; CC: Compensation crystal; TM: Triangular mirror; PPKTP: Periodically-poled Potassium titanyl phosphate; SPAD: Single photon avalanche diode.}
\label{expapp}
\end{figure}

Our aim is to measure a violation of the Clauser-Horne-Shimony-Holt (CHSH) inequality \cite{cla69}
\begin{equation}\label{CHSH}
|\mathcal{S}|=|C(\alpha_1,\beta_1)-C(\alpha_1,\beta_2)+C(\alpha_2,\beta_1)+C(\alpha_2,\beta_2)|\leq2\;,
\end{equation}
where $C(\alpha_j,\beta_k)=\langle \hat{\sigma}_z(\alpha_j)\otimes\hat{\sigma}_z(\beta_k) \rangle$, $j,k=1,2$, and $\hat{\sigma}_z(\theta)=U(\theta)\sigma_z U^\dagger(\theta)$, being $U(\theta)=\binom{\cos\theta\;\;\;\sin\theta}{\sin\theta\;\;\;-\cos\theta}$ and $\sigma_z$ the third Pauli matrix.
We do this by generating polarization-entangled photon pairs and measuring them twice in a row by means of weak-interaction-based measurements.
Specifically, both Alice ($A$) and Bob ($B$) perform two sequential weak measurements on the photon polarization, on the bases set, respectively, by the $\alpha_1,\alpha_2$ and $\beta_1,\beta_2$ parameters.
Although, of course, some decoherence is induced on the measured entangled state, in the weak measurement framework the wavefunction does not (fully) collapse, allowing to measure non-commuting observables in sequence on the same quantum state \cite{pia16seq} and, as a consequence, making it possible to extract information on the whole Bell parameter $|\mathcal{S}|$ from each entangled pair measured.\\
In our realization, we generate polarization-entangled photon pairs, on which we implement weak measurements by exploiting birefringence in thin calcite (CaCO$_3$) crystals.
The experimental setup is presented in Fig. \ref{expapp}.
A CW pump laser at $405$ nm enters the Sagnac interferometer, hosting a periodically-poled Potassium titanyl-phosphate (ppKTP) crystal in which degenerate collinear Type-II spontaneous parametric down conversion (SPDC) occurs, creating orthogonally-polarized photon pairs at $810$ nm.
The L$_P1$ and L$_P2$ lenses are placed to obtain a collimated Gaussian beam inside the ppKTP crystal.
An optical isolator (OI) avoids possible back-reflections to the laser source and selects the laser light to be in a linear polarization, whose axis is eventually rotated by a half-wave plate (HWP$_P$).
A dichroic mirror (DM$_{B}$) sends the pump both to the clockwise and counterclockwise paths of the Sagnac interferometer, allowing to create polarization-entangled photon pairs.
Dichroic mirrors (DM$_{A(B)}$) stop the pump transmission, but allow the down-converted photons to pass through.
The photon pairs are spectrally filtered, coupled to single-mode fibers and then collimated into Gaussian spatial modes $|f_{x_{A(B)}}\rangle\otimes|f_{y_{A(B)}}\rangle$ (with $\langle\xi|f_{\xi}\rangle=\frac{1}{(2\pi\sigma^2)^{1/4}}\exp\left(-\frac{\xi^2}{4\sigma^2}\right)$, being $\xi=x,y$ and $\sigma$ the Gaussian width in both the $x$ and $y$ directions), addressed to two symmetrical branches $(A,B)$ in which the sequential weak measurements take place.\\
The combination of quarter- and half-wave plates (QWP$_{A(B)}$ and HWP$_{A(B)}$, respectively) is used both to compensate the polarization changes due to the single-mode fibers and to adjust the relative phase within the two-photon entangled state, in order to produce the singlet state $\ket{\psi_-}=(\ket{HV}-\ket{VH})/\sqrt{2}$, being $H(V)$ the horizontal (vertical) polarization component.
This way, we are able to initialize the $\ket{\psi_-}$ state with visibility $V^{\mathrm{in}}=0.986\pm0.001$, clearly highlighting the high quality of the entangled states produced by our Sagnac interferometer.
In the end, the initial bipartite state takes the form $\ket{\Psi_{\mathrm{in}}}=\ket{\psi_-}\otimes\ket{f_{x_A}}\otimes\ket{f_{y_A}}\otimes\ket{f_{x_B}}\otimes\ket{f_{y_B}}$.\\
In both branches $A$ and $B$, the sequential weak measurements needed for evaluating the Bell parameter are implemented by exploiting the weak coupling between polarization and transverse momentum induced by two thin birefringent crystal pairs.
In each pair, the principal crystal (PC$_{A(B)}$) induces a tiny spatial mismatch (on a transverse axis with respect to the photon propagation direction) between the $H$ and $V$ polarization components, also generating temporal and phase delay.
A compensation crystal, CC$_{A(B)}$, is added to recover the temporal delay and restore the proper phase without introducing any further spatial decoherence.
The optical $e$-axes of the two consecutive principal crystals on the same branch lie along perpendicular planes, i.e., $z-x$ and $y-z$, in order to make the two measurements independent.
A pair of half-wave plates in each branch (HWP$_{A(B)}$1 and HWP$_{A(B)}$2) allow performing weak measurements in different polarization bases by means of the entangling unitary transformation:
\begin{equation}\label{coupling}
\hat{U}_{\xi_{jK}}=\exp\left( -\frac{i}{\hbar} g_{\xi_{jK}} \hat{\Pi}({\theta_K}_j) \otimes \hat{P}_{\xi_{jK}} \right),
\end{equation}
where $g_{\xi_{jK}}\ll1$ is the weak coupling constant ($K=A,B$, $j=1,2$, $\xi_1=x$, $\xi_2=y$), $\hat{\Pi}({\theta_K}_j)$ represents the polarization projector along a direction described by the angle $\theta_{K_j}$ ($\theta_A=\alpha$, $\theta_B=\beta$) , and $\hat{P}_{\xi_{jK}}$ is the transverse momentum of the photon on the optical plane of the crystal.
Then, the photons are sent to a $24\times24$ Single Photon Avalanche Diode (SPAD) array, i.e. a single photon detector with 2D spatial resolution recording the arrival time and position of each detected photon \cite{mad21}.
The L1 and L2 lenses constitute an imaging system needed to match the photon spatial distributions dimensions with the ones of the SPAD array active area.\\
Thanks to the internal time-tagger of the SPAD array, it is possible to extract the coincidence counts tensor $N(X_A,Y_A,X_B,Y_B)$,
%
%
being $X_A,Y_A$ and $X_B,Y_B$ the coordinates in which the two photon constituting the entangled pair impinge.
This tensor can be used to evaluate, for each detected pair, the entire Bell parameter $\mathcal{S}$.
In fact, one can demonstrate that, considering only projections onto the real plane of the Bloch sphere, after the entangled state undergoes the four weak interactions one has
%
%
%
\begin{equation}
\langle\hat{\xi}_{jA}\otimes\hat{\xi}_{lB}\rangle_{\mathrm{out}} = \bra{\Psi_{\mathrm{out}}}\hat{\xi}_{jA}\otimes\hat{\xi}_{lB} \ket{\Psi_{\mathrm{out}}}= g_{\xi_{jA}}g_{\xi_{lB}} \bra{\psi_-} \hat{\Pi}(\alpha_j)\otimes\hat{\Pi}(\beta_l) \ket{\psi_-}
\end{equation}
and
\begin{equation}
\langle\hat{\xi}_{jK}\rangle_{\mathrm{out}} =\bra{\Psi_{\mathrm{out}}}\hat{\xi}_{jK}\ket{\Psi_{\mathrm{out}}}= g_{\xi_{jK}}\bra{\psi_-}\hat{\Pi}(\theta_{Kj})\ket{\psi_-},
\end{equation}
where $j,l=1,2$, $\hat{\Pi}(\theta_{Kj})=\frac{I+\sigma_z(\theta_{Kj})}{2}$ and $\ket{\Psi_{\mathrm{out}}}=U_{x_A}U_{x_B}U_{y_A}U_{y_B}\ket{\Psi_{\mathrm{in}}}$ is the bipartite state outgoing the measurement process.
Then, the correlation $C(\alpha_j,\beta_l)$ can be written as:
\begin{equation}
C(\alpha_j,\beta_l)=\langle \hat{\sigma}_z(\alpha_j)\otimes\hat{\sigma}_z(\beta_l) \rangle =4\frac{\langle \hat{\xi}_{jA}\otimes\hat{\xi}_{lB} \rangle}{g_{\xi_{jA}}g_{\xi_{lB}}} -2\frac{\langle \hat{\xi}_{jA}\rangle}{g_{\xi_{jA}}} -2 \frac{\langle \hat{\xi}_{jB}\rangle}{g_{\xi_{jB}}} +1,
\end{equation}
and, from that, the Bell parameter $\mathcal{S}$ can be evaluated.\\
Eventually, flipping mirrors FM$_{A}$1 and FM$_{B}$1 allow sending the photons to a projective measurement apparatus, comprising, in each branch, a QWP+HWP+PBS set followed by a SPAD working as a bucket detector.
With this alternative configuration, we can fully characterize the state by performing a quantum tomographic reconstruction \cite{bog10} of its density matrix.

\section{Results}

\begin{figure}[htbp]
\begin{center}
\includegraphics[width=0.95\columnwidth]{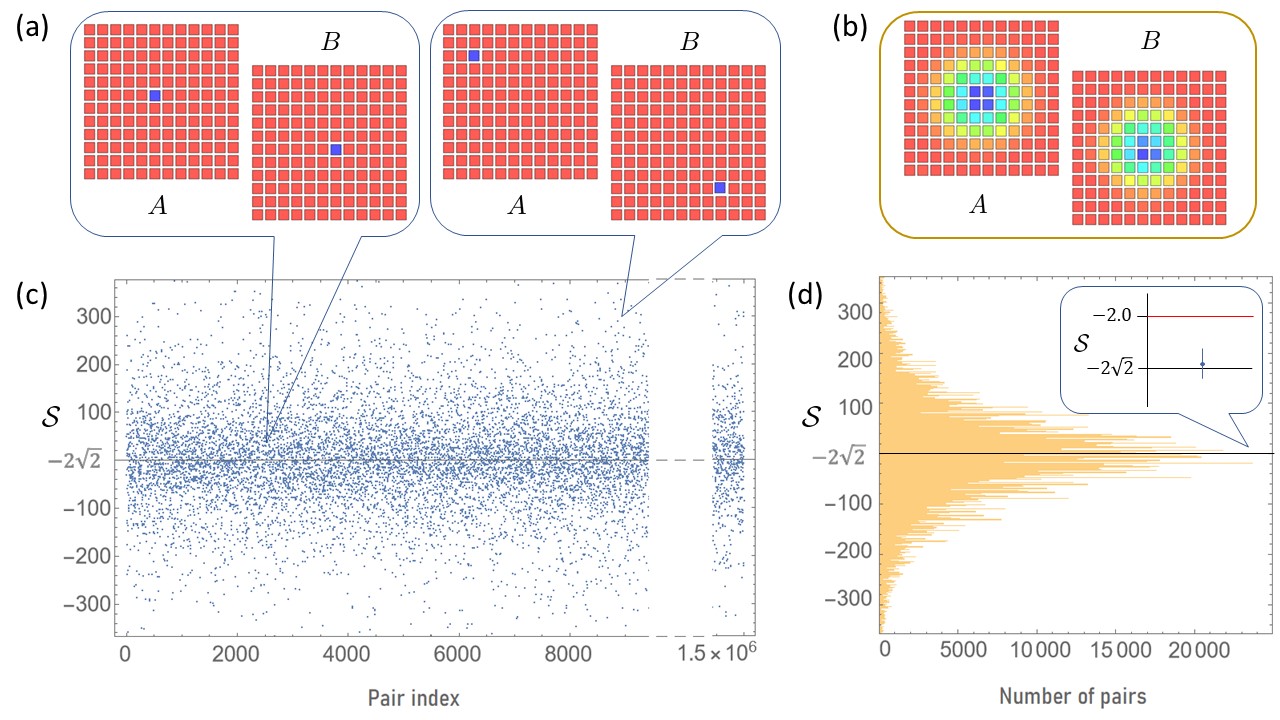}
\caption{Single-pair estimation of the CHSH inequality parameter $\mathcal{S}$. Plot (a): typical entangled pair detection event on the SPAD array. The blue squares indicate the pixels firing in branch $A$ and $B$, respectively. Plot (b): integral of all coincidence count events occurring in both branches. Plot (c): estimation of the $\mathcal{S}$ parameter for each entangled pair detected. Plot (d): histogram representing the integral of all the CHSH parameter estimations. The inset reports the average on all the collected events, $\mathcal{S}^{\mathrm{ave}}=-(2.79\pm0.18)$ (blue dot), in excellent agreement with quantum mechanics Tsirelson bound (black line) ($\mathcal{S}^{\mathrm{QM}}=-2\sqrt2$) and constituting a 4.3 standard deviations violation of the LHVT bound ($\mathcal{S}^{\mathrm{LHVT}}=2$, red line). The uncertainty associated with the $\mathcal{S}^{\mathrm{ave}}$ value includes both statistical and systematic contributions.
} \label{dataset}
\end{center}
\end{figure}

\begin{figure}[htbp]
\begin{center}
\includegraphics[width=0.65\columnwidth]{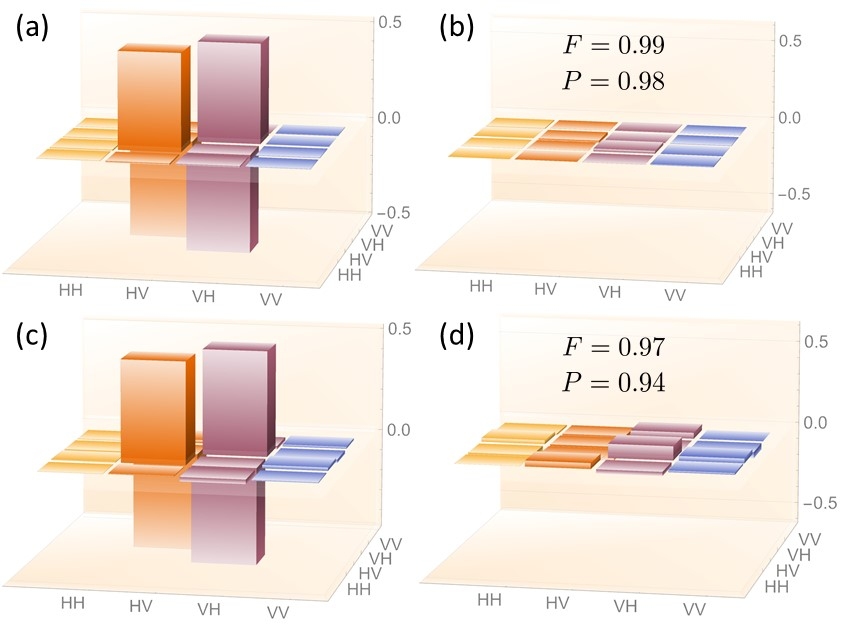}
\caption{Plots (a) and (b), respectively: real and imaginary part of the quantum tomographic reconstruction of the density matrix of the entangled state undergoing the sequential weak measurements needed for the single-pair evaluation of the Bell parameter. The good agreement between the reconstructed density matrix and the theoretically expected one $\rho_{\psi_-}=|\psi_-\rangle\langle\psi_-|$ is certified by the quantum Fidelity $F$ reported in the inset of plot (b), together with the state Purity $P$.
Plots (c) and (d), respectively: real and imaginary part of the quantum tomographic reconstruction of the density matrix of the entangled state after the weak measurement. Again, in the inset of plot (d) we report the Fidelity of the state with the singlet state $\rho_{\psi_-}$, as well as the state Purity.
The high value of both parameters provide evidence of the slight disturbance induced in the entangled state by measurement process instead of the full wavefunction collapse, preserving a large amount of the initial entanglement within the state.}\label{tomo}
\end{center}
\end{figure}

The experimental results are shown in Fig. \ref{dataset}.
There, we show how we are able to estimate the CHSH parameter $\mathcal{S}$ for each of the entangled photon pairs detected.
The huge variance of the results distribution is the signature of the weakness of the measurements occurring in our experiment.
Indeed, weak measurements allow avoiding wavefunction collapse at the cost of obtaining only partial information on the measured observable from a single measurement event, in agreement with the high uncertainty associated with the weak measurement paradigm.
When averaging over multiple detection events, we obtain the significant Bell parameter estimation $\mathcal{S}^{\mathrm{ave}}=-(2.79\pm0.18)$, in perfect agreement with the Tsirelson bound and highlighting a 4.3 standard deviations violation of the classical bound pertaining to LHVTs \cite{gen05}.\\
In addition, to check the state outgoing the measurement process, we first verify the visibility of the state after the Bell parameter (weak) measurement, obtaining $V^{\mathrm{out}}=0.941\pm0.001$, a clear hint of the presence of a large amount of residual entanglement at the end of the protocol.\\
Then, we proceed to the tomographic reconstruction of the state, obtaining the results reported in Fig. \ref{tomo}, where we compare the reconstructed density matrices of the state before [plots (a) and (b)] and after [plots (c) and (d)] the sequential weak measurements.
In both cases, we report as inset the quantum Fidelity $F(\rho^{\mathrm{rec}}, \rho_{\psi_-})=({\rm Tr}(\sqrt{\sqrt{\rho^{\mathrm{rec}}}\rho_{\psi_-}\sqrt{\rho^{\mathrm{rec}}}}))^2$ \cite{gil05} between the reconstructed density matrix $\rho^{\mathrm{rec}}$ and the singlet state one $\rho_{\psi_-}=|\psi_-\rangle\langle\psi_-|$, as well as the Purity of the reconstructed state $P=\mathrm{Tr}\left[(\rho^{\mathrm{rec}})^2\right]$.
The values reported in Fig. \ref{tomo}(b) highlight the high quality of the singlet state produced by our entangled photon source.
The ones shown in Fig. \ref{tomo}(d), instead, allow appreciating how the state outgoing the single-pair Bell parameter measurement is still close to the initial state, a clear sign of the tiny decoherence induced by the four weak measurements (instead of the full collapse of the wavefunction) and, as a consequence, of the large amount of entanglement still present within the state after the measurement process.\\
To give a quantitative estimate of it, we evaluate the Negativity and Concurrence \cite{ver01} of the entangled state entering ($\mathcal{N}^{\mathrm{in}}$ and $\mathcal{C}^{\mathrm{in}}$, respectively) and going out of ($\mathcal{N}^{\mathrm{out}}$ and $\mathcal{C}^{\mathrm{out}}$, respectively) the Bell measurement.
To do this, we follow two different methods: 1) we evaluate them directly from the reconstructed density matrices, obtaining, respectively, $\mathcal{N}^{\mathrm{in}}=0.981$ and $\mathcal{C}^{\mathrm{in}}=0.979$ for the initial state, and $\mathcal{N}^{\mathrm{out}}=0.937$ and $\mathcal{C}^{\mathrm{out}}=0.894$ for the final one; 2) we exploit some optimal estimators \cite{bri10,bri11,vir19} suited for this family of entangled states, achieving, without having to implement the full quantum state tomography procedure, $\mathcal{N}^{\mathrm{in}}=\mathcal{C}^{\mathrm{in}}=0.983\pm0.001$ and $\mathcal{N}^{\mathrm{out}}=\mathcal{C}^{\mathrm{out}}=0.927\pm0.001$.
Aside from the good agreement between the outcomes of these two estimation methods, the extracted Negativity and Concurrence values show how the vast majority of the entanglement initially present in each photon pair is preserved during the Bell measurement, granting the possibility to exploit such a quantum resource for further quantum information protocols or quantum foundations-related tests.

\section{Conclusions}

We have realized the first experiment able to extract information on the whole CHSH parameter from each entangled pair detected, achieving what could be called a single-pair Bell parameter measurement, a real paradigm shift in this framework.
This is obtained by implementing two sequential weak measurements on both particles forming the entangled state under test, granting unprecedented measurement capability and allowing to perform on the same entangled state all the (non-commuting) measurements needed for the Bell inequality evaluation.
Since there is no need to choose among different measurement bases, this protocol is intrinsically unaffected by the freedom-of-choice loophole, paving the way, in principle, for easier implementations of loophole-free tests.
Although the very nature of weak measurements leads to a large uncertainty on a single measurement event (with the exception of the protocol in Ref. \cite{reb21}, that anyway cannot be implemented in this kind of scenario), when averaging on several events a strong violation of the classical bound is obtained.\\
It is also noteworthy that, since the state does not collapse after the (weak) measurements for the Bell inequality, leaving as their mark a slightly decohered entangled state instead, one might even consider utilizing the residual entanglement as a quantum resource for further protocols in quantum information, quantum metrology and related quantum technologies.
In conclusion, our result represents a unicum in the Bell inequality tests framework, not only paving the way to a new generation of experiments addressed to quantum foundations investigation, but also representing a tool for quantum technologies, in particular for non-destructive entanglement tests.

%
%
%
%

\section*{Acknowledgements}
This work was financially supported by the project QuaFuPhy (call ``Trapezio'' of Fondazione San Paolo), by the European Union Horizon 2020 Research and Innovation Programme under FET-OPEN Grant Agreement No. 828946 (PATHOS), by the Israel Innovation Authority under grants 70002 and 73795, by the Elta Systems Ltd., the Pazy Foundation, the Israeli Ministry of Science and Technology, and by the Quantum Science and Technology Program of the Israeli Council of Higher Education.
This work was also funded by the project EMPIR 19NRM06 METISQ.
This project received funding by the EMPIR program cofinanced by the Participating States and from the European Union Horizon 2020 Research and Innovation Programme.
The results presented in this paper have been achieved also in the context of Project EQUO (European QUantum ecOsystems), which is funded by the European Commission in the Digital Europe Programme under the grant agreement No 101091561.
We thank Yakir Aharonov and Avshalom Elitzur for enlightening discussions, and Federico Maestri for contributing to the initial implementation of the experimental setup.



%


\begin{thebibliography}{50}
%
\bibitem{bel65} Bell, J.S. \emph{Physics} 1, 195 (1965).

\bibitem{eke91} Ekert, A. K. Quantum cryptography based on Bell's theorem. \emph{Phys. Rev. Lett.} 67, 661 (1991).

\bibitem{gen05} Genovese, M. Research on hidden variable theories: A review of recent progresses. \emph{Phys. Rep.} 413, 319-396 (2005).

\bibitem{bru13} Brunner, N., Cavalcanti, D., Pironio, S., Scarani, V. \& Wehner, S. Bell nonlocality. \emph{Rev. Mod. Phys.} 86, 419 (2013).

\bibitem{hen15} Hensen, B. et al. Loophole-free Bell inequality violation using electron spins separated by 1.3 kilometres. \emph{Nature} 526, 682 (2015).

\bibitem{giu15} Giustina, M. et al., Significant-Loophole-Free Test of Bell's Theorem with Entangled Photons. \emph{Phys. Rev. Lett.} 115, 250401 (2015).

\bibitem{sha15} Shalm, L.K. Strong Loophole-Free Test of Local Realism. \emph{Phys. Rev. Lett.} 115, 250402 (2015).

\bibitem{fre72} Freedman, J.S. \& Clauser, J.F. \emph{Phys. Rev. Lett.} 28, 938 (1972).

\bibitem{kas75} Kasday, L.R., Ullman, J.D., Wu, C.S. \emph{Nuovo Cimento B} 25, 633 (1975).

\bibitem{cla76} Clauser, J.F. \emph{Phys. Rev. Lett.} 36, 1223 (1976).

\bibitem{asp82} Aspect, A., Dalibard, J. \& Roger, G. Experimental Test of Bell's Inequalities Using Time-Varying Analyzers. Phys. Rev. Lett. 49 (1982) 1804.

\bibitem{aha88} Aharonov, Y., Albert, D.Z. \& Vaidman, L. How the Result of a Measurement of a Component of the Spin of a Spin-1/2 Particle Can Turn Out To Be 100. \emph{Phys. Rev. Lett.} 60, 1351 (1988).

\bibitem{rit91} Ritchie, N.W.M., Story, J.G. \& Hulet, R.G. Realization of a Measurement of a Weak Value. \emph{Phys. Rev. Lett.} 66, 1107 (1991).

\bibitem{kof12} Kofman, A.G., Ashhab, S. and Nori, F. Nonperturbative theory of weak pre-and post-selected measurements. \emph{Phys. Rep.} 520, 43 (2012).

\bibitem{tam13} Tamir, B. \& Cohen, E. Introduction to Weak Measurements and Weak Values. \emph{Quanta} 3, 7 (2013).

\bibitem{mit07} Mitchison, G., Jozsa, R., and Popescu, S. Sequential weak measurement. \emph{Phys. Rev. A} 76, 062105 (2007).

\bibitem{the16} Thekkadath, G.S., Giner, L., Chalich, Y., Horton, M.J., Banker, J. \& and Lundeen, J.S. Direct Measurement of the Density Matrix of a Quantum System. \emph{Phys. Rev. Lett.} 117, 120401 (2016).

\bibitem{pia16seq} Piacentini, F. et al. Measuring Incompatible Observables by Exploiting Sequential Weak Values. \emph{Phys. Rev. Lett.} 117, 120402 (2016).

\bibitem{kim18} Kim, Y., Kim, YS., Lee, SY., Moon, S., Kim, YH. \& Cho, YW. Direct quantum process tomography via measuring sequential weak values of incompatible observables. \emph{Nat. Commun.} 9, 192 (2018).

\bibitem{fol21} Foletto, G., Padovan, M., Avesani, M., Tebyanian, H., Villoresi, P. \& Vallone, G. Experimental Test of Sequential Weak Measurements for Certified Quantum Randomness Extraction. \emph{Phys. Rev. A} 103, 062206 (2021).

\bibitem{pia16pus} Piacentini, F. et al. Experiment Investigating the Connection between Weak Values and Contextuality. \emph{Phys. Rev. Lett.} 116, 180401 (2016).

\bibitem{wae17} Waegell, M., Denkmayr, T., Geppert, H., Ebner, D., Jenke, T., Hasegawa, Y., Sponar, S., Dressel, J. \& Tollaksen, J. Confined contextuality in neutron interferometry: Observing the quantum pigeonhole effect. \emph{Phys. Rev. A} 96, 052131 (2017).

\bibitem{cim20} Cimini, V., Gianani, I., Piacentini, F., Degiovanni, I.P. \& Barbieri, M. Anomalous values, Fisher information, and contextuality, in generalized quantum measurements. \emph{Quantum Sci. Technol.} 5, 025007 (2020).

\bibitem{mah16} Mahler, D.H., Rozema, L., Fisher, K., Vermeyden, L., Resch, K.J., Wiseman, H.M. \& Steinberg, A. Experimental nonlocal and surreal Bohmian trajectories. \emph{Sci. Adv.} 2, e150146 (2016).

\bibitem{hu18} Hu MJ., Zhou, ZY., Hu, XM., Li, CF., Guo, GC. \& Zhang, YS. Observation of non-locality sharing among three observers with one entangled pair via optimal weak measurement. \emph{npj Quantum inf.} 4, 63 (2018).

\bibitem{gog11} Goggin, M.E., Almeida, P.M., Barbieri, M., Lanyon, B.P., O'Brien, J.L., White, A.G. \& Pryde, G.J. Violation of the Leggett-Garg inequality with weak measurements of photons. \emph{PNAS} 108, 1256 (2011).

\bibitem{ave17} Avella, A. et al., Anomalous weak values and the violation of a multiple-measurement Leggett-Garg inequality. \emph{Phys. Rev. A} 96, 052123 (2017).

\bibitem{cal20} Calder\'{o}n-Losada, O., Moctezuma Quistian, T.T., Cruz-Ramirez, H., Ramirez, S. M., U'Ren, A.B., Botero, A. \& Valencia, A. A weak values approach for testing simultaneous Einstein-Podolsky-Rosen elements of reality for non-commuting observables. \emph{Commun. Phys.} 3, 117 (2020).

\bibitem{mad21} Madonini, F., Severini, F., Incoronato, A., Conca, E. \& Villa, F. Design of a 24x24 SPAD imager for multi-photon coincidence-detection in super resolution microscopy. Proc. of SPIE 11771, Quantum Optics and Photon Counting 2021, 117710B (2021).

\bibitem{cla69}  Clauser, J.F., Horne, M.A., Shimony, A. \& Holt, R.A. Proposed experiment to test local hidden-variable theories. \emph{Phys. Rev. Lett.} 23, 880  (1969).

\bibitem{bog10} Bogdanov, Y. U., Brida, G., Genovese, M., Kulik, S.P., Moreva, E.V. \& Shurupov, A.P. Statistical Estimation of the Efficiency of Quantum State Tomography Protocols \emph{Phys. Rev. Lett.} 105, 010404 (2010).

\bibitem{gil05} Gilchrist, A., Langford, N.K. \& Nielsen, M.A. Distance measures to compare real and ideal quantum processes. \emph{Phys. Rev. A} 71, 062310 (2005).

\bibitem{ver01} Verstraete, F., Audenaert, K., Dehaene, J. \& De Moor, B. A comparison of the entanglement measures negativity and concurrence. \emph{J.
Phys. A: Math. Gen.} 34, 10327 (2001).

\bibitem{bri10} Brida, G. et al. Experimental estimation of entanglement at the quantum limit. \emph{Phys. Rev. Lett.} 104, 100501 (2010).

\bibitem{bri11} Brida, G. et al. Optimal estimation of entanglement in optical qubit systems. \emph{Phys. Rev. A} 83, 052301 (2011).

\bibitem{vir19} Virz\`{i}, S., Rebufello, E., Avella, A., Piacentini, F., Gramegna, M., Ruo Berchera, I. Degiovanni, I.P. \& Genovese, M. Optimal estimation of entanglement and discord in two-qubit states. \emph{Sci. Rep.} 9, 3030 (2019).

\bibitem{reb21} Rebufello, E. et al. Anomalous weak values via a single photon detection. \emph{Light: Sci. \& Appl.} 10, 106 (2021).










\end{thebibliography}
\end{document}